\documentclass[pmlr]{jmlr}

\usepackage{longtable}

\usepackage{booktabs}
\usepackage[load-configurations=version-1]{siunitx} 
\usepackage{breakcites}

 \usepackage{tikz}
 \usetikzlibrary{automata, positioning, arrows}
 \usepackage{amsmath}
 \usepackage{amsfonts}
 \usepackage{xfrac}    
 \usepackage{booktabs}
 \usepackage{float}

 \newcommand{\NP}{\textsf{NP}}

 \newcommand{\DFACONS}{\textsc{DFA-Con}\ }

 \newcommand{\DFACONSspace}{\textsc{DFA-Con}}

 \DeclareMathOperator{\lang}{\mathcal{L}}

 \newcommand{\posSet}{\ensuremath{P}}
 \newcommand{\negSet}{\ensuremath{N}}
 \usepackage{comment}
 \excludecomment{cancel}
 \excludecomment{excludetoappendix}
 \usepackage{todonotes}

 \usepackage{algorithm}
 \usepackage{algorithmic}

\theorembodyfont{\upshape}
\theoremheaderfont{\scshape}
\theorempostheader{:}
\theoremsep{\newline}

\jmlrworkshop{LearnAut 2022}

\title[Learning from Positive and Negative Examples]{Learning from Positive and Negative Examples: \\New Proof for Binary Alphabets}

  \author{\Name{Jonas Lingg} \addr Eberhard Karls Universit{\"a}t T{\"u}bingen, Germany \Email{j.lingg@mailbox.org} \\
   \Name{Mateus de Oliveira Oliveira\nametag{\thanks{Supported by the RCN projects 288761 and 326537.}}} 
   \addr University of Bergen, Norway \Email{mateus.oliveira@uib.no} \\
\Name{Petra Wolf\nametag{\thanks{Supported by DFG project FE 560/9-1.}}} 
\addr Universit{\"a}t Trier, Germany \Email{wolfp@informatik.uni-trier.de}}

\begin{document}

\maketitle
\begin{abstract} One of the most fundamental problems in computational learning
theory is the problem of learning a finite automaton $A$ consistent with a
finite set $\posSet$ of positive examples and with a finite set $\negSet$ of
negative examples. By consistency, we mean that $A$ accepts all strings in
$\posSet$ and rejects all strings in $\negSet$. It is well known that this
problem is \NP-complete. In the literature, it is stated that \NP-hardness
holds even in the case of a binary alphabet. As a standard reference for this
theorem, the work of Gold from 1978 is either cited or adapted. But as a
crucial detail, the work of Gold actually considered Mealy machines and not
deterministic finite state automata (DFAs) as they are considered nowadays. As
Mealy automata are equipped with an output function, they can be more compact
than DFAs which accept the same language.  We show that the adaptations of
Gold's construction for Mealy machines stated in the literature have some
issues, and provide a correct proof for the fact that the {\em DFA-consistency
problem} for binary alphabets is NP-complete.  
\end{abstract}
\begin{keywords}
Learning Automata, Computational Complexity, Finite Samples
\end{keywords}

\section{Introduction}
\label{sec:intro}

In the {\em DFA-consistency problem} (\DFACONSspace) we are given 
a pair of disjoint sets of strings $\posSet,\negSet\subseteq \Sigma^*$
and
a positive integer $k$. The goal is to determine whether there is a deterministic finite automaton (DFA) $A$ with at most $k$
states that accepts all strings in $\posSet$ and rejects all strings in $\negSet$. 
Even though~\cite{DBLP:journals/jacm/PittW93} showed that the problem cannot be approximated within any polynomial factor,
\DFACONSspace~has become one of the most central problems in computational learning theory (see~\cite{DBLP:journals/iandc/Gold67,DBLP:journals/iandc/Angluin78,DBLP:journals/iandc/Gold78,DBLP:conf/aii/Pitt89,parekh2001learning})
with applications that span several subfields of artificial intelligence and related areas,
such as automated synthesis of controllers \cite{ramadge1987supervisory}, model checking \cite{groce2002adaptive,mao2016learning} 
optimization \cite{najim1990optimization,yazidi2020team,bouhmala2015multilevel,coste2006learning}  
neural networks \cite{meybodi2002new,hasanzadeh2018learning,mayr2018regular,guo2019learning}, multi-agent systems~\cite{nowe2005learning}, 
and others  \cite{rezvanian2018recent,najim2014learning,de2010grammatical}. 


The \DFACONS problem has been studied for at least five decades, including parameterized analysis by~\cite{DBLP:journals/jcss/FernauHV15}, and it was stated multiple times that the \NP-completeness of the \DFACONS problem also holds in the case of binary alphabets, see~\cite{DBLP:journals/iandc/Angluin78,de_la_higuera_2010,DBLP:journals/jcss/FernauHV15}.
In~\cite{DBLP:journals/iandc/Angluin78}, the work of~\cite{DBLP:journals/iandc/Gold78} is cited for this fact and in~\cite{de_la_higuera_2010} and \cite{DBLP:journals/jcss/FernauHV15} adaptations of the construction from~\cite{DBLP:journals/iandc/Gold78} are given.
The issue is that \NP-hardness results for variants of 
\DFACONS considered in \cite{DBLP:journals/iandc/Gold78} are actually stated in terms of Mealy machines, and do not translate directly to the modern context of DFAs in the case of binary alphabets. By taking a closer look on the preliminaries, \cite{DBLP:journals/iandc/Angluin78} is also considering Mealy machines.
For the adaptations in \cite{de_la_higuera_2010} and~\cite{DBLP:journals/jcss/FernauHV15}, both of them follow the 
same approach, and both contain inaccuracies that invalidate the proofs.
They have in common that they are adaptations of the construction by Gold for the consistency problem of Mealy machines~\cite{DBLP:journals/iandc/Gold78}
but as the Mealy machines considered in~\cite{DBLP:journals/iandc/Gold78} (mapping $\Sigma^*\to \Gamma$, $|\Gamma| = 2$) can be more compact (when interpreted as language acceptors) than DFAs recognizing the same language, the difference in number of states causes the adaptations to fail. 

We solve this issue by giving a new construction for the claim that the \DFACONS problem is \NP-complete when restricted to binary alphabets. 

This work is structured as follows. After giving necessary definitions, we present our main result, a new construction for the \NP-hardness of \DFACONS restricted to binary alphabets. Then, we discuss in more detail why the reductions in~\cite{de_la_higuera_2010,DBLP:journals/jcss/FernauHV15} fail and why the construction of Gold does not directly transfer to the setting of DFAs.

We thank an anonymous reviewer of LearnAut 2022 for suggesting a simpler construction than we had initially.


\section{Preliminaries}
For a finite alphabet $\Sigma$, we call $\Sigma^*$ the set of all words over
$\Sigma$. For a
regular expression~$r$ and a fixed number~$k$, we denote $k$ concatenations of
$r$ with $r^k$.  A \textit{deterministic finite automaton} (DFA) is a tuple
$A=(Q,\Sigma,\delta,s_0,F)$ where $Q$ is a finite set of states, $\Sigma$ a
finite alphabet, $\delta~\colon~Q \times \Sigma \rightarrow Q$ a total
transition function, $s_0$ the initial state and $F \subseteq Q$ the set of
final states. We call $A$ a \emph{complete} DFA if we want to highlight that
$\delta$ is a \emph{total} function. If $\delta$ is partial, we call $A$ a
partial DFA.
We generalize $\delta$ to words by $\delta(q, aw) = \delta(\delta(q, a), w)$
for $q \in Q, a\in \Sigma, w \in \Sigma^*$. We further generalize $\delta$ to
sets of input letters $\Gamma\subseteq \Sigma$ by $\delta(q, \Gamma) =
\bigcup_{\gamma\in\Gamma} \{\delta(q, \gamma)\}$. A DFA $A$ accepts a word
$w\in\Sigma^*$ if and only if $\delta(s_0, w) \in F$. We let $\lang(A)$ denote
the {\em language of $A$}, i.e., the set of all words accepted by $A$. 
We denote the empty word with $\epsilon$, $|\epsilon|=0$. 
The \DFACONS problem is formally defined as follows.

\begin{definition}[\DFACONSspace]
	\ \\
	Input: Finite set of words $P, N \subseteq \Sigma^*$ with $P \cap N = \emptyset$, and integer $k$.\\
	Question: Is there a DFA $A = (Q, \Sigma, \delta, s_0, F)$ with $|Q| \leq k$, $P \subseteq \lang(A)$, and $\lang(A) \cap N = \emptyset$?
\end{definition}

\section{Main Result}
As our main result, we present an adaptation for DFAs of the reduction by Gold.
\begin{theorem}
	\label{thm:WAA-Sigma-1}
	Restricted to binary alphabets, \DFACONS is \NP-complete.
\end{theorem}

	We show that there is a reduction 
    from \textsc{All-pos-neg 3SAT}\footnote{Given a Boolean formula in conjunctive normal form where all clauses have at most three literals (this form is called 3CNF) and all literals in a clause are either all positive or all negative. Is there a variable assignment satisfying the formula?} to \DFACONS mapping each instance of \textsc{All-pos-neg 3SAT} with $n \geq 1$ variables and $m \geq 1$ clauses to an instance of \DFACONS with $k = n + m$, $|\Sigma| = 2$, $|\posSet| = \mathcal{O}(m)$, and $|\negSet| = \mathcal{O}(m^2+nm)$ in time $\mathcal{O}(n^3+n^2m)$.
	
    Let $\phi$ be a Boolean formula in 3CNF with $n \geq 1$ variables
$$V = \{x_0, x_1, \dots, x_{n-1}\}$$ and $m\geq 1$ clauses $C = \{c_0, c_1, \dots, c_{m-1}\}$, where each clause contains either only positive literals or only negative literals.
Given $\phi$, we let $k = n + m$ and 
    \begin{align*}
        \posSet = &\{\epsilon, a^k\} \cup \{a^{i}bb \mid i \in [0, \ldots, m - 1], \text{$C_{i}$ positive} \}\\
        \negSet = &\{a^i\mid 0 < i < k\} \\ 
                  &\cup \{a^{i}bb \mid i \in [0, \ldots, m - 1], \text{$C_{i}$ negative}\}\\
                  &\cup \{a^{i}ba^{k-r} \mid (0 \leq i \leq  m - 1) \land (0 \leq r < k) \land (r < m \lor x_{r - m } \notin C_{i})\}
    \end{align*}
    be the corresponding instance of \DFACONSspace.

    We show that there exists a satisfying variable assignment $$\beta \colon V \to \{\texttt{false}, \texttt{true}\}$$ if and only if there exists a DFA with at most $k$ states that is consistent with $\posSet$ and $\negSet$.

\begin{figure}[h!]
        \centering
        \begin{tikzpicture}
        	[->,>=stealth',shorten >=1pt,auto,node distance=1.5cm,
        	semithick,state/.style={circle, draw}]
        	\node[state,initial,accepting,fill=blue!20] (0) at (0,0) {};
        	\node[state, right of=0,fill=blue!20] (1) {};
        	\node[state, right of=1,fill=blue!20] (2) {};
        	\node[state, right of=2,fill=pink!40] (3) {};
        	\node[state, right of=3,fill=pink!40] (4) {};
        	\node[state, right of=4,fill=pink!40] (5) {};
        	\node[above of=1] {\textcolor{blue}{clauses}};
        	\node[above of=4] {\textcolor{pink}{variables}};
        	
        	\path
        	(0) edge node {$a$} (1)
        	(1) edge node {$a$} (2)
        	(2) edge node {$a$} (3)
        	(3) edge node {$a$} (4)
        	(4) edge node {$a$} (5)
        	(1) edge[bend left] node {$b$} (4)
        	(2) edge[bend left] node {$b$} (5)
        	(0) edge[bend left] node {$b$} (3)
        	(4) edge[bend left=20] node {$b$} (0)
        	(5) edge[bend left=40] node {$a,b$} (0)
        	(3) edge[loop below] node {$b$} (3);
        \end{tikzpicture}
        \caption{Automaton for $\phi = \neg x_0 \wedge x_1 \wedge x_2$.}%
        \label{fig:aut_reduction}
    \end{figure}
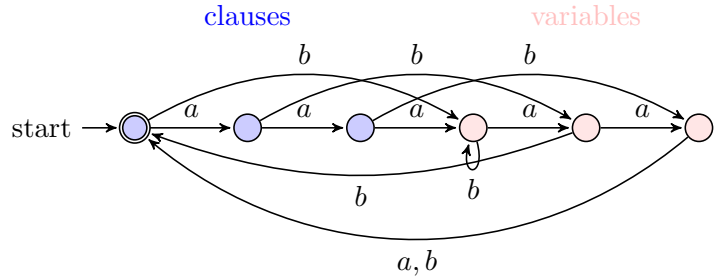

    We start with a remark on the structure of a DFA $A$ with at most $k$ states accepting $\{\epsilon, a^k\}$ and
    rejecting $\{a^i\mid 0 < i < k\}$. 
    Let $s_i$ be the state of $A$ reached after reading $a^i$. For each two distinct $i$ and $j$ in $\{0,1,\dots,k-1\}$, 
$A$ accepts after reading $a^{k-i}$ from $s_i$, but rejects after reading $a^{k-j}$ from $s_i$. Therefore, 
    the states $s_0,\dots,s_{k-1}$ are pairwise distinct. This implies that $s_k = s_0$, and that $A$ consists of 
    a loop of $k$ states $s_0,s_1,\dots,s_{k-1}$   where $s_0$ is the sole initial and the sole accepting state, and for each $i\in \{0,1,\dots,k-1\}$,
  $s_i$ transitions to $s_{i+1\mod k}$ when reading letter $a$
   (see \autoref{fig:aut_reduction}). 

    The intuition is that the first $m$ states, $s_0, s_1, s_{m-1}$, correspond to the clauses
    $C_0, C_2, \ldots, C_{m-1}$ and the $n$ last states, $s_m, s_{m+1}, \ldots, s_{m+n-1}$,
    correspond to the variables $x_0, x_1, \ldots, x_{n-1}$. 
 
    We start by showing that if $\phi$ is satisfiable, then there exists a DFA $A$ with $k$ states consistent with $P$ and $N$.  Let $\beta \colon V \to \{\texttt{false}, \texttt{true}\}$ be
    a satisfying variable assignment. We let $A = (\{s_0, s_1, \ldots, s_{k-1}\},$ $\{a,b\}, \delta, s_0, \{s_0\})$.
     For each $i\in \{0,1,\dots,k-1\}$, we let $\delta(s_i, a) = s_{i+1 \mod k}$. The transitions corresponding to letter $b$ depend 
    on the satisfying assignment $\beta$. For each $i\in \{0,1,\dots,m-1\}$, let $x_j$ be a variable in $C_i$ such that its assignment under $\beta$ satisfies $C_i$. 
     Note that since $\beta$ is a satisfying assignment, such variable always exists.  
     We set 
    $\delta(s_{i}, b) = s_{m+j}$. Intuitively, if the automaton is at the state $s_i$ corresponding to clause $C_i$, then after reading symbol $b$ 
    it transitions to the state $s_{m+j}$ corresponding to variable $x_j$. 
    For each $j\in \{0,1,\dots,n-1\}$, if $\beta(x_j) = \texttt{true}$, then we set  $\delta(s_{m+j}, b) = s_0$. Otherwise, 
    we set $\delta(s_{m+j}, b) = s_{m+j}$. Intuitively, if the automaton is at the state $s_{m+j}$ corresponding to variable $x_j$, then after reading 
symbol $b$, it transitions to the accepting state $s_0$ if $x_j$ is set to 
\texttt{true} and to the state $s_{m+j}$ otherwise (note that $s_{m+j}$ is a rejecting state).
    See \autoref{fig:aut_reduction} for an example.

    Now, we show that $A$ accepts all words in $\posSet$ and rejects all words in $\negSet$.
    By construction, $A$ accepts $\{\epsilon, a^k\}$ and rejects $\{a^i\mid 0 < i < k\}$.
    Let $a^{i}bb \in \{a^{i}bb \mid i \in [0, \ldots, m - 1], \text{$C_{i}$ positive} \}$. 
    After reading $a^i$, $A$ transitions to state $s_i$ corresponding to the positive clause $C_{i}$.
    By construction, reading $b$ from $s_i$ leads to the state $s_{m+j}$ corresponding to a variable
    $x_j$ contained in $C_i$ such that $\beta(x_j) = \texttt{true}$. Therefore,
    $\delta(\delta(s_i,b),b) = s_0$ and $A$ accepts $a^{i}bb$. Similarly, $A$
    rejects all the words in
    $\{a^{i}bb \mid i \in [0, \ldots, m - 1], \text{$C_{i}$ negative} \}$. Finally,
    $A$ rejects all the words in the set 
    \begin{equation}
    \label{equation:NegativeSamples}
    \{a^{i}ba^{k-r} \mid (0 \leq i \leq m - 1) \land (0 \leq r < k) \land (r < m \lor x_{r - m} \notin C_{i})\}. 
    \end{equation}
    To see this, note that for each $i\in \{0,\dots,m-1\}$, the state reached by the string $a^ib$ is the state $s_{m+j}$ corresponding to some variable 
    $x_j$ in $C_i$. Therefore, given $r\in \{0,1,\dots,k-1\}$, the string $a^iba^{k-r}$ is accepted if and only if $r=m+j$. 
    Nevertheless, this is never the case, since either $r< m$, or $m\leq r < k$ and $x_{r-m} \notin C_i$. 

    For the converse, suppose that there exists a DFA $A$ with at most $k$ states consistent with $P$ and
    $N$. As we saw already, $A$ has exactly $k$ states $s_0, s_1, \ldots, s_{k-1}$
    that form a loop with the letter $a$, and $s_0$ is both the initial state and the unique accepting state.
    For each $i\in \{0,1,\dots,m-1\}$, the negative samples in (\ref{equation:NegativeSamples})
    ensure that the transition with $b$ from state $s_{i}$ points to the state $s_{m+j}$ corresponding to 
    a variable $x_j$ contained in  $C_i$, since if this were not the case, some string in the set $(\ref{equation:NegativeSamples})$ would have been accepted by $A$. 
    Now, we define an assignment $\beta$ of variables depending on the transitions labelled
    with $b$. For each $j\in \{0,1,\dots,n-1\}$, if the transition labeled by $b$ going from state $s_{m+j}$ points to the accepting state $s_0$, then 
we set $\beta(x_j) = \texttt{true}$. Otherwise, we set $\beta(x_j) = \texttt{false}$.
    For each $i\in \{0,1,\dots,m-1\}$, let $C_i$ be a positive clause in $\phi$. By consistency, $A$ accepts the word $a^{i}bb$. 
    Note that after reading string $a^{i}$, $A$ reaches the state $s_{i}$ corresponding to $C_i$, and then, 
    by reading $b$ the automaton transitions to the state $s_{m+j}$ corresponding to some variable $x_j$ contained in $C_i$. 
    Finally, after reading the last $b$ the automaton transitions to the unique accepting state $s_0$. Therefore, by construction, 
    $\beta(x_j) = \texttt{true}$ and the assignment of $x_j$ satisfies $C_i$.
    A similar argument works if $C_i$ is a negative clause. Therefore $\beta$ satisfies
    $\phi$. $\square$

\section{Counterexamples Regarding \DFACONS with $|\Sigma| =2$} 
\label{sec:counterex}
In the literature, several proofs for the \NP-hardness of \DFACONS for fixed alphabet size of $|\Sigma|=2$ have appeared. We give in the following a counterexample which contradict the proofs in~\cite{DBLP:journals/jcss/FernauHV15} and in~\cite{de_la_higuera_2010}. We further discuss in this section why the claim does not follow directly from the construction in~\cite{DBLP:journals/iandc/Gold78}. 
\subsection*{Counterexample (\cite{DBLP:journals/jcss/FernauHV15})}
\begin{figure}[h!]
    \centering
    \begin{tikzpicture}[->,>=stealth',shorten >= 1pt,auto,node distance = 1.5cm,semithick,scale=0.75, every node/.style={transform shape}] 
      \node[state, initial, accepting] (1) {$t$};
      \node[state] (2) [right of = 1] {$xy$};
      \node[state] (3) [right of = 2] {$yz$};
      \node[state] (4) [right of = 3] {$vx$};
      \node[state] (5) [right of = 4] {$\bar{x}\bar{z}$};
      \node[state] (6) [right of = 5] {$\bar{v}\bar{y}$};
      \node[state] (7) [right of = 6] {$\bar{x}\bar{y}$};
      \node[state] (8) [right of = 7] {$x$};
      \node[state] (9) [right of = 8] {$y$};
      \node[state] (10) [below of = 9] {$z$};
      \node[state] (11) [below of = 10] {$v$};
      \node[state] (12) [below of = 11] {$f$};
  
      \draw
      (1) edge [loop above] node{$b$} (1)
      (1) edge node{$a$} (2)
      (2) edge node{$a$} (3)
      (2) edge[bend right=-45] node[above]{$b$} (8)
      (3) edge node{$a$} (4)
      (3) edge[bend right=-40] node{$b$} (8) 
      (4) edge node{$a$} (5)
      (4) edge[bend right=-35] node{$b$} (8)
      (5) edge node{$a$} (6)
      (5) edge[bend right=60] node[below]{$b$} (9)
      (6) edge node{$a$} (7)
      (6) edge[bend right=50] node{$b$} (9)
      (7) edge node{$a$} (8)
      (7) edge[bend right=48] node{$b$} (9)
      (8) edge node{$a$} (9)
      (8) edge[bend left=-60] node[above]{$b$} (1)
      (9) edge node{$a$} (10)
      (9) edge[bend right=-30] node{$b$} (12)
      (10) edge node{$a$} (11)
      (11) edge node{$a$} (12)
      (12) edge[bend left] node{$a,b$} (1)
      (10) edge[loop left] node{$b$} (10)
      (11) edge[loop left] node{$b$} (11)
      ;
    \end{tikzpicture}
    \caption{Automaton corresponding to $\phi$.}
    \label{fig:cnt1}
\end{figure}
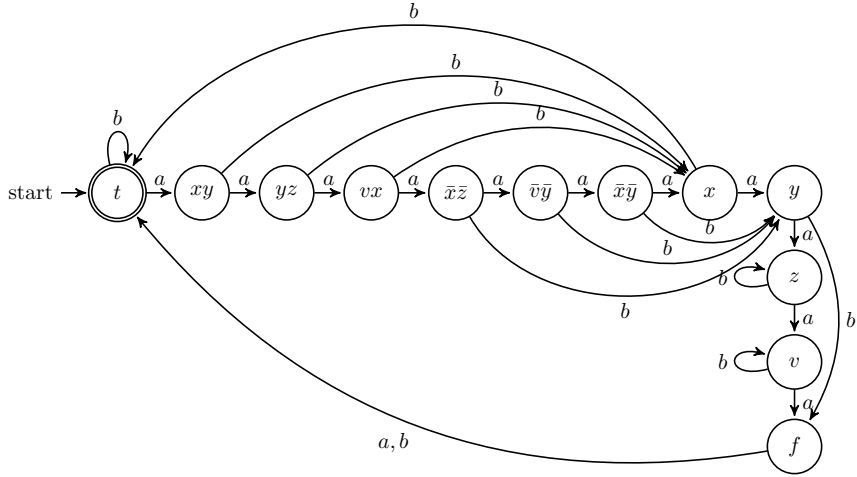 

\noindent We give a Boolean formula that is unsatisfiable and shows that the construction used in the proof of Lemma 15 in \cite{DBLP:journals/jcss/FernauHV15} allows us to construct an automaton that is consistent with $\posSet$ and $\negSet$. Consider the unsatisfiable formula
\[\phi=(x \lor y) \land (y \lor z) \land (v \lor x) \land (\lnot x \lor \lnot z) \land (\lnot v \lor \lnot y) \land (\lnot x \lor \lnot y)\] with $n=4$ variables and $m=6$ clauses. We index the clauses from left to right in the formula above, starting with $i=1$. We construct the \DFACONS instance $\posSet$, $\negSet \subseteq \{a, b\}^*$, $k=m+n+2=12$ according to the construction presented in the proof of Lemma~15 in~\cite{DBLP:journals/jcss/FernauHV15}:
%
\begin{align*}
	\posSet=\, &\{\epsilon, b, a^{12}, a^{11}b\}\cup \{a^ibbb\mid1 \leq i \leq 6\} \cup \{a^ibb\mid 1 \leq i \leq 3\} \cup \{a^ibba \mid 4 \leq i \leq 6\},\\
	\negSet=\, &\{a^k\mid1 \leq k \leq 11\} \cup \{a^ib,a^iba\mid 1 \leq i \leq 6\} \cup \{a^ibb \mid 4 \leq i \leq 6\}.
\end{align*}

Figure \ref{fig:cnt1} shows a DFA with 12 states over $\{a, b\}$ that is consistent with $\posSet$ and $\negSet$ despite the formula $\phi$ being unsatisfiable. This refutes Lemma 15 in \cite{DBLP:journals/jcss/FernauHV15}.

\subsection*{Counterexample (\cite{de_la_higuera_2010})}
\label{apx:counter2}
\begin{figure}[h!]
    \centering
    \begin{tikzpicture}[->,>=stealth',shorten >= 1pt,auto,node distance = 2cm,semithick,scale=0.75, every node/.style={transform shape}] 
\node[state, initial, accepting] (1) {$1$};
\node[state] (2) [right of = 1] {$2$};
\node[state] (3) [right of = 2] {$3$};
\node[state] (4) [right of = 3] {$4$};
\node[state] (5) [right of = 4] {$5$};
\node[state] (6) [right of = 5] {$6$};
\node[state] (7) [right of = 6] {$7$};
\node[state] (8) [right of = 7] {$8$};
\node[state] (9) [above of = 4] {};
      \draw

      (1) edge node{$a$} (2)
      (1) edge[bend left=15] node{$b$} (9)
      (2) edge node{$a$} (3)
      (2) edge node{$b$} (9)
      (3) edge node{$a$} (4)
      (3) edge[bend right=55] node{$b$} (8)
      (4) edge node{$a$} (5)
      (4) edge[bend right=50] node{$b$} (8)
      (5) edge node{$a$} (6)
      (5) edge[bend right=40] node{$b$} (8)
      (6) edge node{$a$} (7)
      (6) edge node{$b$} (9)
      (7) edge node{$a$} (8)
      (7) edge[bend right=15] node{$b$} (9)
      (8) edge[bend left=60] node{$a,b$} (1)
      (9) edge[loop above] node{$a, b$} (9) 
      ;
    \end{tikzpicture}
    \caption{Automaton with $n+1=9$ states corresponding to the clause set $\{\{\lnot{x_2},\lnot{x_3},\lnot{x_5}\},$ $\{\lnot{x_8}\},\{x_1,x_4,x_8\},\{x_1,x_6,x_8\},\{x_6,x_7,x_8\},\{\lnot{x_4},\lnot{x_6}\},\{\lnot{x_1},\lnot{x_6}\},\{\lnot{x_1},\lnot{x_7}\}\}$.}
    \label{fig:cnt2}
\end{figure}
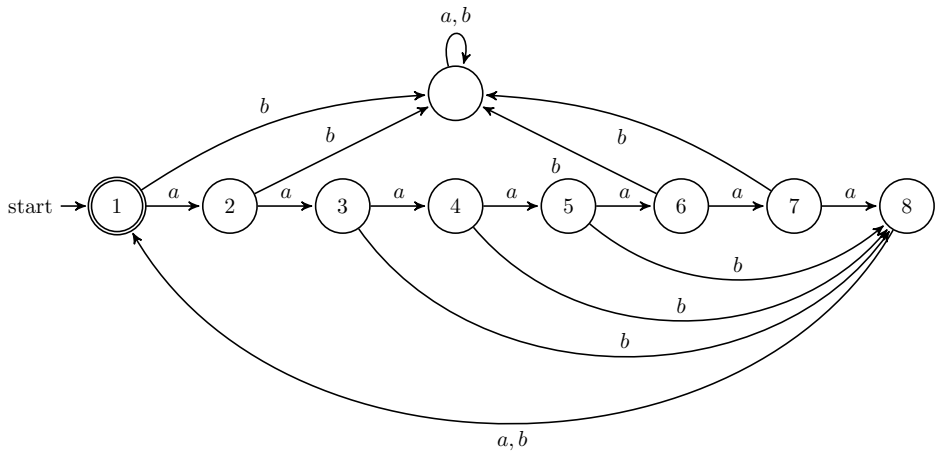

The set of clauses 
$C = \{\{\lnot{x_2},\lnot{x_3},\lnot{x_5}\}_1,\{\lnot{x_8}\}_2,\{x_1,x_4,x_8\}_3,\{x_1,x_6,x_8\}_4,$ $\{x_6,x_7,x_8\}_5,$ $\{\lnot{x_4},\lnot{x_6}\}_6,\{\lnot{x_1},\lnot{x_6}\}_7,\{\lnot{x_1},\lnot{x_7}\}_8\}$ 
represents an \emph{unsatisfiable} Boolean formula $\phi$ with $n=8$ variables and $m=8$ where each clause is either purely positive or purely negative. The index imposes an ordering on the clauses. The variable-set 
$V=\{x_1,x_2,x_3,x_4,x_5,x_6,x_7,x_8\}$
is also ordered. In accordance with the construction in the proof of Theorem $6.2.1$ in~\cite{de_la_higuera_2010}, page 120, we define the sets
\begin{align*}
	\posSet=&\{a^8\} \cup \{a^{i-1}ba^8b \mid i \in \{3,4,5\}\},\\
	\negSet=&\{a^k, a^{8+k} \mid 1 \leq k < 8\} \cup \{a^{i-1}ba^8b \mid i \in \{1,2,6,7,8\}\}\\
	&\cup \{a^{i-1}ba^{8-j+1} \mid i \in \{3,4,5\}, x_j \notin C_i\} \\ 
        &\cup \{a^{i-1}ba^{8-j+1} \mid i \in \{1,2,6,7,8\}, \lnot x_j \notin C_i\}.
\end{align*}
using the two orderings that are implicit in the proof for Theorem $6.2.1$ in the textbook. Claim~1 in the proof of Theorem $6.2.1$ claims that every DFA consistent with $\posSet$ and $\negSet$ has at least $n+1$ states. In contrast, Claim~3 speaks about the existence of an $n$-state DFA. The precise state-bound of the constructed instance in the proof of Theorem $6.2.1$ is not explicitly given. We show that the construction with the state-bound given in Claim~1 cannot be correct by giving an $(n+1)$-state DFA in Figure \ref{fig:cnt2}, which is consistent with the sets $\posSet$ and $\negSet$ obtained from an unsatisfiable formula.
Also for a bound of exact $n$ states, this proof is not correct, as the example discussed with respect to Gold's construction below in this section is also a counterexample for the construction of Theorem $6.2.1$ in~\cite{de_la_higuera_2010}.
Consider the formula $\varphi = \neg x_1 \wedge x_2 \wedge x_3$ with clause set $\{\{\neg x_1\}, \{x_2\}, \{x_3\}\}$ and variable set $\{x_1, x_2, x_3\}$ implicitly ordered from left to right. Here, $m=3$ and $n=3$.
This formula is clearly satisfiable and has pure clauses, but there is no DFA with~3 states that can separate the two sets of words of the de la Higuera construction properly.
Let's call the three states $q_1$, $q_2$, $q_3$. Since $a^3 \in \posSet$ and $\{a^1, a^2, a^4, a^5\} \subseteq \negSet$ we get for $a$: 
$\delta(q_1, a) = q_2$, $\delta(q_2, a) = q_3$, $\delta(q_3, a) = q_1$ and $q_1$ must be an accepting state.
Now, consider the transition with $b$ for the state $q_2$. 
Since the second clause $\{x_2\}$ $(i=2)$ is positive we have by the third item of the construction $a^{i-1}ba^nb \in \posSet$, hence $a^1ba^3b \in \posSet$.
By the second condition of this item we have for $x_3 \notin \{x_2\} \colon a^{i-1}ba^{n-j+1} \in \negSet$, hence $a^1ba^1 \in \negSet$. From this point we also get for $x_1 \notin \{x_2\} \colon a^1ba^3 \in \negSet$.
These three words now lead to a contradiction concerning $\delta(q_2, b)$.
We cannot have $\delta(q_2, b) = q_3$ since then $\delta(q_1, aba) = q_1$, which contradicts $aba \in \negSet$.
We cannot have $\delta(q_2, b) = q_1$ since then $\delta(q_1, abaaa) = q_1$, which contradicts $abaaa \in \negSet$.
Hence, we must have $\delta(q_2, b) = q_2$. But then, $\delta(q_1, a^1ba^3b) = q_2$, which contradicts $abaaab \in \posSet$.
Hence, we cannot find a three-state DFA consistent with the construction. 

\subsection*{General Issue}
Both of the constructions (\cite{DBLP:journals/jcss/FernauHV15} and \cite{de_la_higuera_2010}) rely on the construction in the proof of Theorem 2 in~\cite{DBLP:journals/iandc/Gold78} where a reduction from the `Satisfiability Question' to the problem `Is there a finite automaton with states reachable by $T$ which agrees with $D$?' is given. Here, $T$ is a finite set of words representing states, i.e., the words in $T$ are elements of the equivalence classes representing states of the sought automaton; moreover no two elements of $T$ are in the same class. The finite set $D$ (set of data) consists of pairs of words over the input alphabet and single output letters. The key point is that in~\cite{DBLP:journals/iandc/Gold78} the term automaton refers to `Mealy model finite state automaton' and hence, the automata do not include a set of final states, instead they are enhanced with an output function, which can be interpreted as indicating the acceptance or rejection of a word. For a binary output alphabet, these two models are equivalent considering the class of recognizable languages, but the number of states needed to recognize a language may differ. For instance, the language $\{0,1\}^*0$ can be accepted with a one-state Mealy automaton but for DFAs it can only be accepted by a two-state DFA. 
Coming back to the consistency problem for DFAs, one is tempted to think that a proof for the \NP-hardness for \DFACONS is hidden in the corollary following Theorem 2 in~\cite{DBLP:journals/iandc/Gold78} (`$Q_\text{min}(D,n)$ is \NP-complete for Card($U$) = Card($Y$) = 2'; $U$ is the input alphabet and $Y$ the output alphabet and $Q_\text{min}(D,n)$ asks, given data $D$ and positive integer $n$, is there a finite automaton with $n$ states which agrees with $D$?) but with the arguments above, this is not the case. 
For instance, we can give a counterexample to this claim.
Consider the satisfiable formula $\varphi_1= \neg x_1 \wedge x_2 \wedge x_3$ with $n=3$ and implicit variable order $x_1 < x_2 < x_3$ as well as clause order $\{\neg x_1\} < \{x_2\} < \{x_3\}$. Following the construction in Theorem~2 in~\cite{DBLP:journals/iandc/Gold78}, one can verify that there is a three-state Mealy machine recognizing the words in the state characterization matrix correctly, but there is no three-state DFA consistent with the data. 
Another counterexample is the satisfiable formula $\varphi_2 = x_1 \wedge x_3 \wedge \neg x_2$ with variable order $x_1 < x_3 < x_2$ and clause order $\{x_1\} < \{x_3\} < \{\neg x_2\}$.

\paragraph{Acknowledgements:} Petra Wolf was supported by DFG project FE 560/9-1, and Mateus de Oliveira Oliveira
by the RCN projects 288761 and 326537. 
We thank an anonymous reviewer of LearnAut 2022 for suggesting a simpler construction than we had initially.

\bibliography{bib}

\end{document}